\documentclass[aps,apl,reprint,superscriptaddress]{revtex4-1}
\usepackage{graphicx}
\usepackage{dcolumn}
\usepackage{bm}
\usepackage{color}
\usepackage{sidecap}

\begin{document}

\title{Field-free Superconducting Diode Effect in FeTe$_{0.55}$Se$_{0.45}$}

\author{Peng Dong} 
\affiliation{ShanghaiTech Laboratory for Topological Physics \& School of Physical Science and Technology, ShanghaiTech University, Shanghai 201210, China}
\affiliation{State Key Laboratory of Quantum Functional Materials, ShanghaiTech University, Shanghai 201210, China}

\author{Jinghui Wang}
\affiliation{ShanghaiTech Laboratory for Topological Physics \& School of Physical Science and Technology, ShanghaiTech University, Shanghai 201210, China}
\affiliation{State Key Laboratory of Quantum Functional Materials, ShanghaiTech University, Shanghai 201210, China}

\author{Yanjiang Wang}
\affiliation{ShanghaiTech Laboratory for Topological Physics \& School of Physical Science and Technology, ShanghaiTech University, Shanghai 201210, China}
\affiliation{State Key Laboratory of Quantum Functional Materials, ShanghaiTech University, Shanghai 201210, China}

\author{Jianjun Xiao}
\affiliation{ShanghaiTech Laboratory for Topological Physics \& School of Physical Science and Technology, ShanghaiTech University, Shanghai 201210, China}
\affiliation{State Key Laboratory of Quantum Functional Materials, ShanghaiTech University, Shanghai 201210, China}

\author{Xiang Zhou}
\affiliation{ShanghaiTech Laboratory for Topological Physics \& School of Physical Science and Technology, ShanghaiTech University, Shanghai 201210, China}
\affiliation{State Key Laboratory of Quantum Functional Materials, ShanghaiTech University, Shanghai 201210, China}

\author{Hui Xing}
\affiliation{Key Laboratory of Artificial Structures and Quantum Control and Shanghai Center for Complex Physics, School of Physics and Astronomy, Shanghai Jiao Tong University, Shanghai 200240, China}

\author{Yueshen Wu} \email{wuysh@shanghaitech.edu.cn}
\affiliation{ShanghaiTech Laboratory for Topological Physics \& School of Physical Science and Technology, ShanghaiTech University, Shanghai 201210, China}
\affiliation{State Key Laboratory of Quantum Functional Materials, ShanghaiTech University, Shanghai 201210, China}

\author{Yulin Chen}
\affiliation{ShanghaiTech Laboratory for Topological Physics \& School of Physical Science and Technology, ShanghaiTech University, Shanghai 201210, China}
\affiliation{State Key Laboratory of Quantum Functional Materials, ShanghaiTech University, Shanghai 201210, China}
\affiliation{Department of Physics, Clarendon Laboratory, University of Oxford, Oxford OX1 3PU, United Kingdom}

\author{Jinsheng Wen} \email{jwen@nju.edu.cn}
\affiliation{Collaborative Innovation Center of Advanced Microstructures, Nanjing, 210093, China.}

\author{Jun Li}
\email{lijun3@shanghaitech.edu.cn}
\affiliation{ShanghaiTech Laboratory for Topological Physics \& School of Physical Science and Technology, ShanghaiTech University, Shanghai 201210, China}
\affiliation{State Key Laboratory of Quantum Functional Materials, ShanghaiTech University, Shanghai 201210, China}

\date{\today}

\begin{abstract}

The superconducting diode effect (SDE)—the asymmetry of critical currents with respect to current direction—is a pivotal advancement in non-reciprocal superconductivity. While SDE has been realized in diverse systems, a fundamental challenge remains achieving field-free operation in iron-based superconductors with simple device geometries. Here, we report a non-volatile, field-free SDE in thin crystalline FeTe$_{0.55}$Se$_{0.45}$(FTS), showing asymmetric critical currents with a rectification coefficient of 1.9\% and operating temperatures up to 9 K. Intriguingly, a pronounced non-zero second harmonic resistance emerges at the superconducting transition, exhibiting a sign reversal under varying current and temperature. The SDE persists at zero magnetic field and the rectification coefficient($\eta$) exhibits an even symmetric dependence on the magnetic field, distinguishing it from magnetic chirality anisotropy mechanisms. In addition to this, we systematically ruled out influences from dynamic superconducting domains, thermal gradients, and sample geometry, while establishing that localized stress amplifies the rectification coefficient, likely constituting one of the principal contributing factors. These results establish FTS as a robust platform for realizing field-free superconducting diodes in a structurally simple platform. 

\end{abstract}

\maketitle

\section{Introduction}

The superconducting diode effect (SDE) arises from nonreciprocal transport in superconductors and has become a central topic in condensed matter physics, which manifests as nonreciprocity, where the positive and negative critical currents ($I_{c}^{+}$) and ($I_{c}^{-}$) differ \cite{MaoY,GhoshS,Nadeem,JiangK,YuanN,ZhangY,BauriedlL,LyuYY}. The undirectional transport arises due to noncentrosymmetric Fermi surfaces through explicit symmetry-breaking mechanisms, such as crystal lattice and device structure inducing inversion symmetry breaking, external perturbations like magnetic fields and dynamic superconducting order domains inducing time-reversal symmetry-breaking \cite{Narita,WuH,GhoshS,WanZ,LeT}. The distinctive characteristics inherent of SDE, especially field-free SDE, are highly promising for the realm of superconducting physics such as unconventional superconductors \cite{Bastian} and its applications in electronic device engineering, including superconducting electronics, spintronics, and quantum information \cite{MH}. To date, the field-free SDE has been reported across diverse superconducting systems, including  superconducting superlattice films \cite{Narita}, asymmetric Josephson junction \cite{WuH,GhoshS}, the superconducting order domain \cite{WanZ,LeT}. Importantly, the pursuit of a system exhibiting intrinsic breaking of time-reversal and inversion symmetries serves as an exemplary platform for investigating field-free superconducting diode phenomena.

\begin{figure*}
	\includegraphics[width=0.9\textwidth]{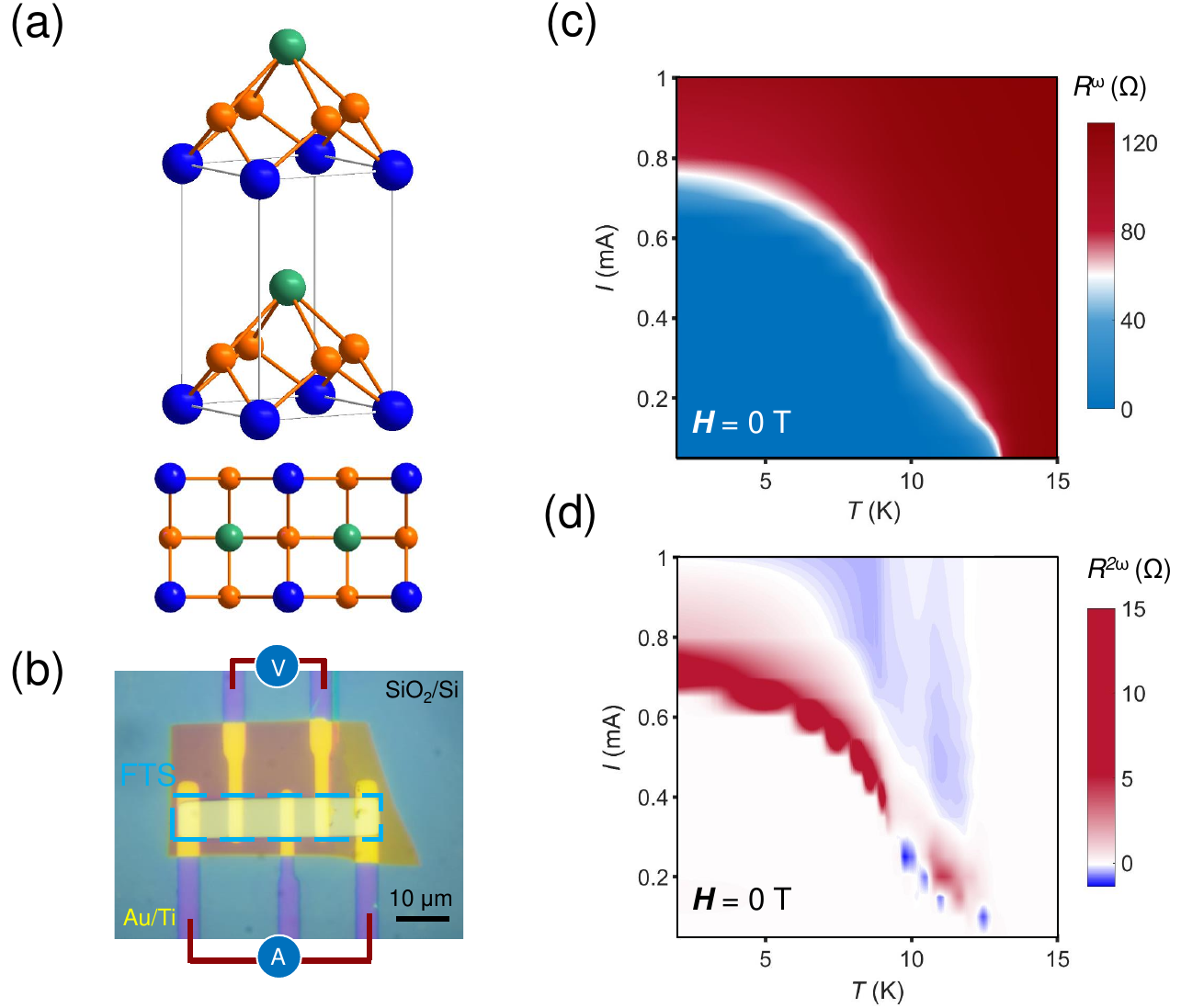}
	\caption{\label{Fig.1} Nonreciprocal transport properties in FeTe\textsubscript{0.55}Se\textsubscript{0.45}(FTS) under zero field (\textbf{\emph{B}} = 0 T) . (a) Crystal structure of FTS inversion symmetry is broken along the c axis. Orange denotes Fe atoms, green signifies Te atoms, and blue represents Se atoms. (b) Optical image of FTS sample. AC or DC current is administered through the electrode positioned at the farthest point to ensure uniform distribution, while the central planar section is designated as the voltage measurement area. In this configuration, the precisely defined rectangular structure comprises the FTS encapsulated with boron nitride(BN), where the violet hue arises from the combined effect of Au/Ti electrodes and SiO\textsubscript{2} etched to a specific thickness. (c) The temperature dependence of first harmonic resistance($R_{\text{xx}}^{\omega}$) under zero field cooling by applying different current range from 50 $\mu$A to 1 mA. Zero-resistance state is attained, at $T_{c}^{0}$ = 13 K under low current excitation in device A. (d) The coresponding second harmonic resistance($R_{\text{xx}}^{2\omega}$) as a function of temperature using AC measurement. Obvious non-zero signal is developed at the superconducting transition and the sign of the resistance even reverses under low current excitation in device A.}
\end{figure*}

Recently, the iron-based superconductor FeTe\textsubscript{0.55}Se\textsubscript{0.45}(FTS) has emerged as a prominent candidate for intrinsic topological superconductivity (TSC) owing to the combination of bulk superconductivity, topological Dirac surface states, and ferromagnetic fluctuations. Crucially, experiments employing angle-resolved photoemission spectroscopy (ARPES) \cite{Peng}, spin-polarized tunneling \cite{Dongfei,Machida}, nitrogen-vacancy magnetometry \cite{McLaughlin}, surface Kerr effect \cite{Farhang}, nano-SQUID imaging \cite{LinYS}, and Josephson junction transport \cite{QiuG} have consistently revealed time-reversal symmetry breaking and support strong evidence of unconventional superconductor. These findings align with observations of Majorana-bound states and robust Yu-Shiba-Rusinov (YSR) surface states in FTS, which underscore its potential for hosting correlated topological superconductivity and Kondo-driven surface dynamics \cite{Chatzopoulos}. FTS is near the critical point of the superconducting, nematic, topological and ferromagnetic states by slightly changing the Te/Se ratio or the interstitial Fe content, which would cause the intricate interplay among these components and their emergent exotic phenomena. The superconducting diode phenomenon, serving as a probe for detecting space and time inversion symmetry breaking, offers a refined experimental lens to investigate such properties within the superconducting phase.


Here, we report the observation of field-free SDE within the iron-based superconductor FTS flakes. Our findings highlight a 1.9\% deviation in critical currents at 2 K. Remarkably, the second harmonic resistance ($R_{\text{xx}}^{2\omega}$) demonstrated a distinct non-zero signal at the superconducting transition, altering its sign depending on the current excitation or temperature variations. Meanwhile, stable half-wave rectification under zero magnetic field conditions have consistently observed. Though, the mirrored reaction to magnetic fields suggests a mechanism distinct from previously noted zero-field SDE. Based on this analysis, we have examined superconducting chiral domains, thermal gradients induced by geometric configurations, and stress-related factors. The results demonstrate a stronger correlation with stress, leading us to attribute the observed phenomenon primarily to stress-induced effects. This field-free SDE phenomenon could streamline the integration of superconducting devices and limit potential interference, thereby offering a promising direction for the development of superconducting electronics utilizing iron-based high-temperature superconductors.

\section{Results and discussion}

   In Fig. 1(a), the crystal structure of FTS is similar to that of FeSe, with inversion symmetry breaking along the $c$-axis. The mirror symmetry may be perturbed along the in-plane direction due to the partial substitution of Se atoms by Te atoms. Owing to the van der Waals bonding between FTS layers, the bulk crystal can be readily exfoliated into atomically thin flakes using conventional all-dry exfoliation and transfer techniques. 
   
\begin{figure*}
	\includegraphics[width=0.9\textwidth]{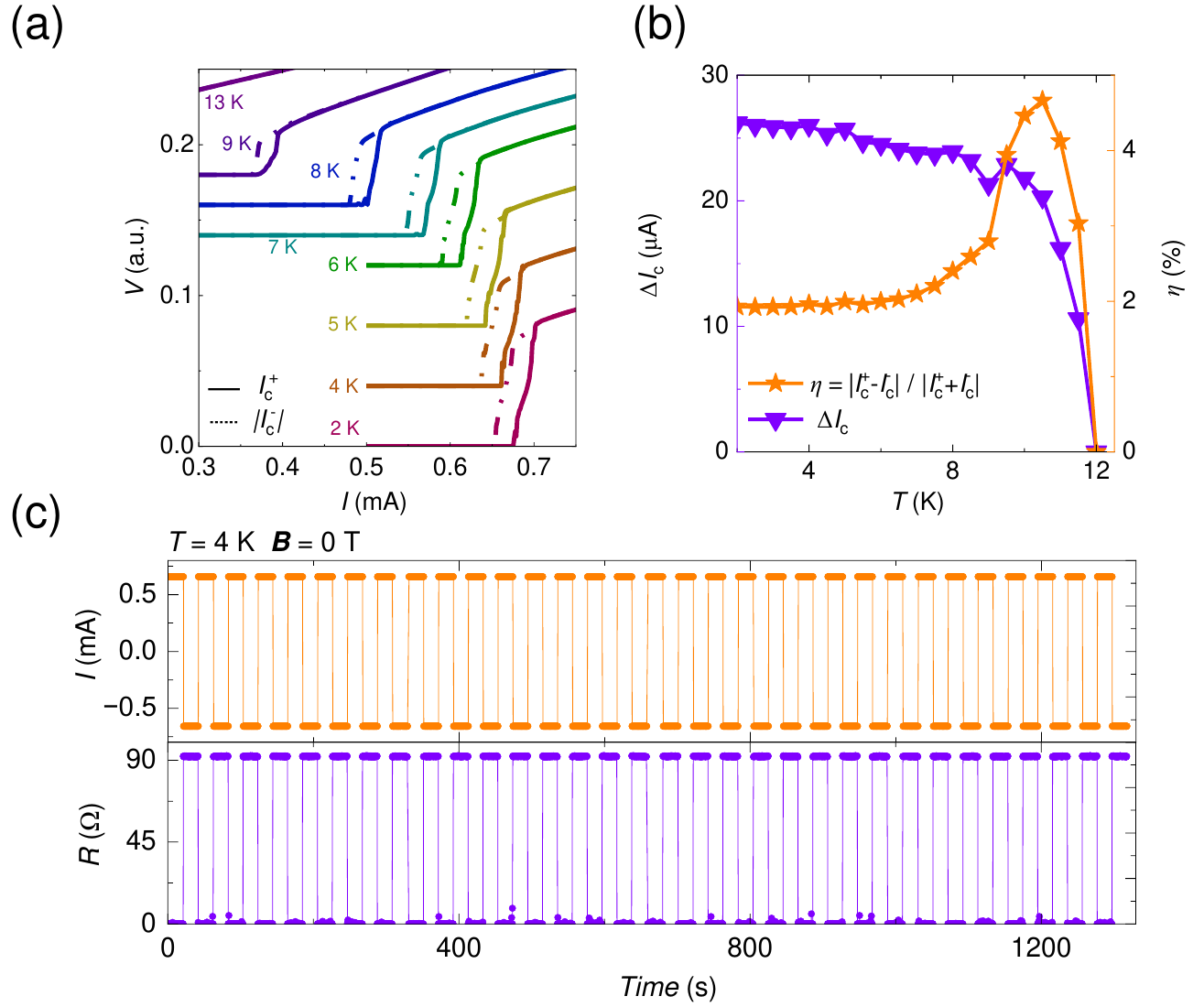}
	\caption{\label{Fig.2} Zero field SDE in Device A. (a) The negative side (from 0 mA to -0.8 mA) and positive side (from 0 mA to 0.8 mA) of current-votage curves at different temperature. A clear difference of critical current is observed. (b) The disparity in critical current (purple line) diminishes progressively with increasing temperature, whereas the rectification coefficient (orange line) remains invariant at low temperature, undergoes initial amplification to 4\% with temperature increasing, and ultimately attenuates to zero. (c) Resistance switch is realized by applying a current of $\pm$0.6575 mA at 4 K and zero magnetic field. The half-wave rectification is achieved with high stability and robustness.}
\end{figure*}

The sample is initially dissociated onto a silicon wafer cleaned via oxygen plasma treatment. Subsequently, a conventional transfer technique employing boron nitride (BN) is utilized to encapsulate the sample. Through a combination of photolithography and etching processes, the specimen is meticulously sculpted into a precisely defined geometry. Further lithographic patterning is then performed on the structured sample, wherein the etching rate of BN at the electrode regions is meticulously regulated until reaching the FTS sample, ensuring optimal ohmic contact formation. An optical image of sample is shown in Fig. 1(b), where the total channel length can reach up to 60~$\mu$m, making any sub-micron physical influence negligible.


  We first characterized the nonreciprocity of superconductivity at zero magnetic field using a standard low-frequency (73~Hz) lock-in technique in a four-probe measurement geometry. All electrical measurements were carried out in a Quantum Design Physical Property Measurement System (PPMS). 
The first and second harmonic resistances were defined as 
$R_{\text{xx}}^{\omega} = \sqrt{2} V_{\text{xx}}^{\omega} / I$ and 
$R_{\text{xx}}^{2\omega} = \sqrt{2} V_{\text{xx}}^{2\omega} / I$, 
where $I$ denotes the amplitude of the sinusoidal AC current, and 
$V_{\text{xx}}^{\omega}$ and $V_{\text{xx}}^{2\omega}$ are the amplitudes of the first and second harmonic voltage signals measured by the lock-in amplifier. A phase shift of $\pi/2$ was applied to the second harmonic component to compensate the phase.

\begin{figure*}
\includegraphics[width=0.9\textwidth]{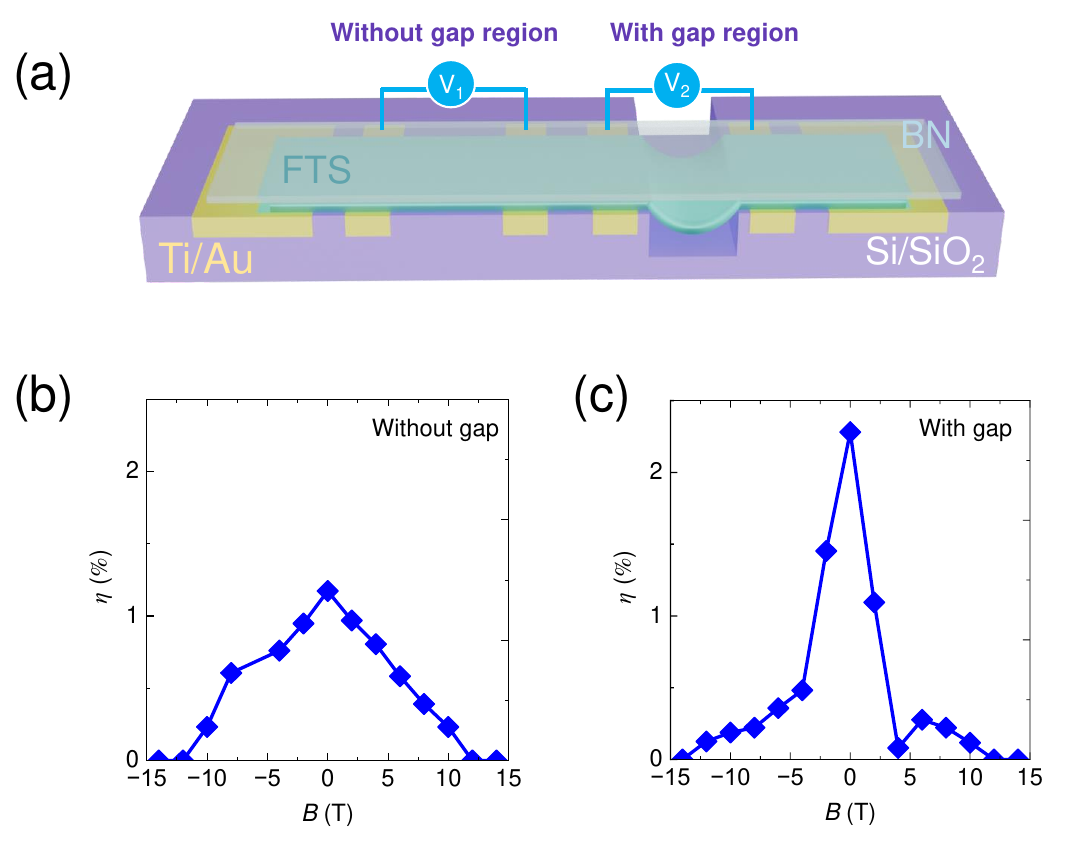} 
\caption{\label{Fig.3} Comparative analysis of rectification coefficients with and without localized stress. (a)Schematic illustration of the sample structure in the stress-application experiment. The right electrode underwent additional photolithography and reactive ion beam etching to create a trench approximately 1.3 $\mu$m deep. Mechanical stress was induced in the sample through compressive deformation during the transfer process. (b)The region devoid of localized stress exhibits an even-symmetry dependence for both rectification coefficient and voltage differential with respect to magnetic field variations. (c)In contrast, the stressed region demonstrates an even-symmetry characteristic in the rectification coefficient, accompanied by a 1\% enhancement, while the voltage differential manifests no discernible dependence on the magnetic field.}
\end{figure*}

  As shown in Fig. 1(c), the critical temperature of device A is approximately 13 K under low current excitation. The superconductivity gradually weakens as larger currents are applied. Interestingly, the second harmonic resistance $R_{\text{xx}}^{2\omega}$ signal becomes non-zero at the superconducting transition, even without the application of magnetic field in Fig. 1(d). Notably, the $R_{\text{xx}}^{2\omega}$ in device A is comparable in magnitude to the first-harmonic resistance $R_{\text{xx}}^{\omega}$, and its sign reverses under low-current excitation, which makes SDE possible. It is important to note that before cooling the device to its superconducting state, the magnet was oscillated to minimize any residual magnetic field to below 1 Oe in a Quantum Design DynaCool system. 

The strong nonreciprocity in device A motivated us to investigate the SDE using direct current (DC) measurements. 
Fig. 2(a) shows the voltage response as a function of DC current at zero magnetic field and various temperatures. Specifically, the solid curve represents the sweep from zero to the positive critical current ($I_{c}^{+}$), while the dashed curve denotes the sweep from zero to the negative critical current ($I_{c}^{-}$). Here, $I_{c}^{+}$ and $I_{c}^{-}$ correspond to the current values at which the device resistance reaches 50\% of its normal-state value.
As temperature decreases, both critical currents evolve accordingly, accompanied by a clear asymmetry between the positive and negative bias directions. At 2~K and zero magnetic field, the critical current in the positive direction reaches 0.693~mA, while that in the negative direction is slightly lower at 0.666~mA.
To rigorously rule out any influence from residual magnetic fields, we performed a 180$^\circ$ rotation of the sample, as schematically illustrated in Fig.~S6. The resulting measurements showed no reversal in the critical current polarity, confirming that remnant magnetic fields have a negligible effect on the observed SDE. This methodological robustness has been further supported by repeated experimental validation~\cite{LeT,QiS}. 

The temperature dependence of the critical current difference between positive and negative bias directions is summarized in Fig.~2(b). A slight decrease in $\Delta I_{c}$ is observed from 2~K to 10~K, followed by a sharp drop to zero at 12~K, indicating the disappearance of the SDE near the superconducting transition temperature.
To quantify the strength of the SDE, we define the rectification coefficient $\eta$ as
\[
\eta = \frac{\left| I_{c}^{+} + I_{c}^{-} \right|}{\left| I_{c}^{+} - I_{c}^{-} \right|} \times 100\%.
\]
As shown in Fig. 2(b), $\eta$ remains nearly constant at approximately 2\% at low temperatures, increases to around 4\% with rising temperature, and then rapidly decreases to 0\% near the critical temperature.
The zero-field SDE was reproducibly observed in 6 devices, with the rectification coefficient $\eta$ ranging from 0.5\% to 4\%, as illustrated in Fig.S3--S8. The rectification factor $\eta$ of the SDE exhibits no apparent dependence on material thickness. To systematically mitigate the influence of Joule heating from current injection, we employed a pulsed excitation protocol. We employed pulsed current excitation to reduce Joule heating and obtain reversible switching (see Fig. S5).

\begin{figure*}
\includegraphics[width=0.9\textwidth]{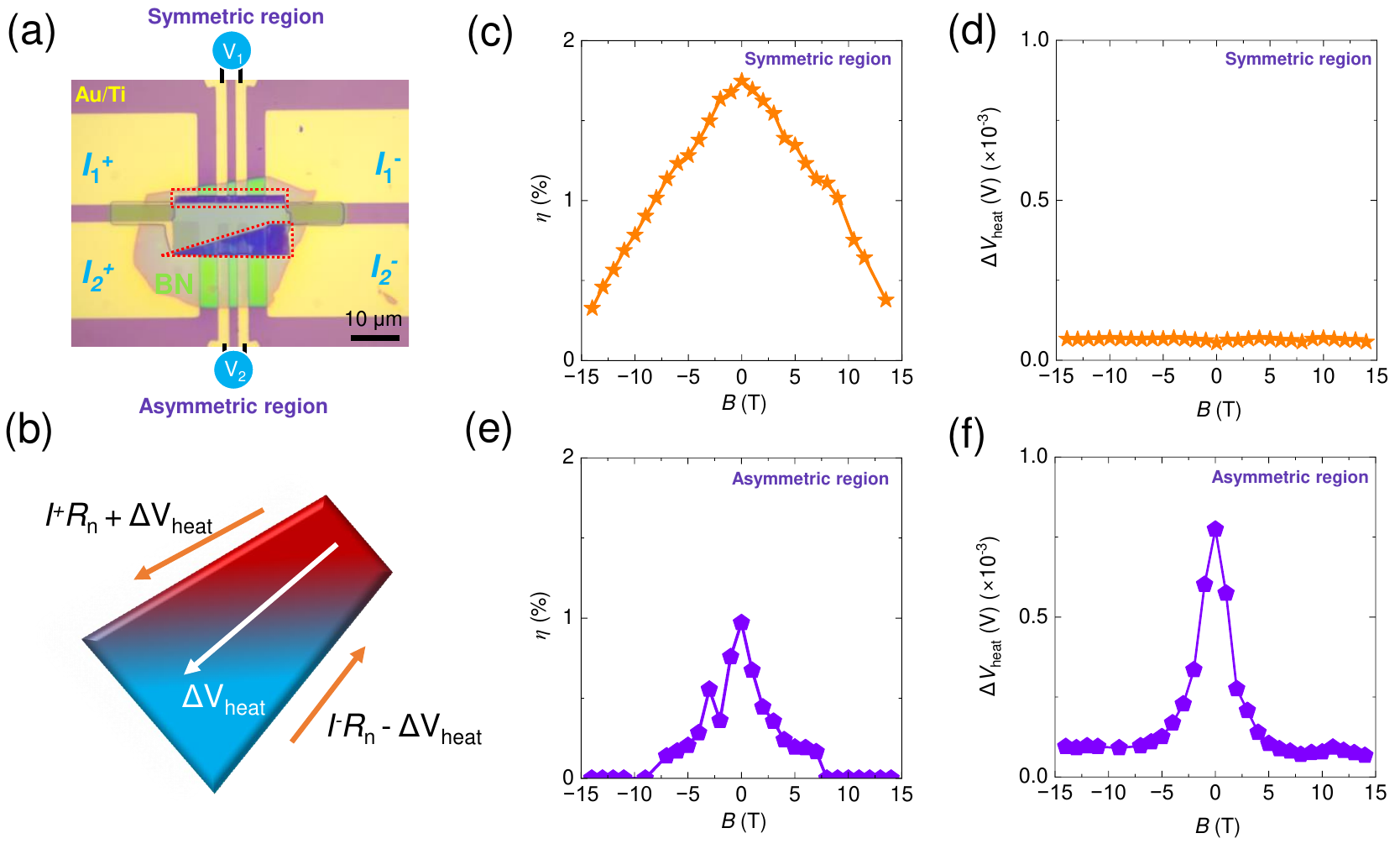} 
\caption{\label{Fig.4} The dependence of rectification coefficient ($\eta$) and thermoelectric potential ($\Delta V_{heat}$) on magnetic fields in symmetric and asymmetric configurations. (a) Optical image of symmetric and asymmetric sample with a scale bar denoting 10 $\mu$m.The device was meticulously transferred via a dry-transfer process onto pre-fabricated electrodes and subsequently encapsulated with boron nitride for protection. This was followed by sophisticated micro-nano fabrication techniques to selectively etch the same sample into both symmetrical and asymmetrical configurations. (b) Illustration of the intrinsic potential voltage difference ($\Delta V_{heat}$) arising from thermal gradient in the specimen.The voltage disparity for equivalent currents of opposite polarity in the normal state manifests as twice the $\Delta V_{heat}$. (c) The rectification coefficient $\eta$ dependence of applied magnetic fields manifests unequivocal even-symmetry. (d) The voltage differential($\Delta V_{heat}$) calculated by $| V_{+0.8 mA} - V_{-0.8 mA}|$ exhibits no discernible dependence on the magnetic field. (e) The rectification coefficient $\eta$ dependence of applied magnetic fields manifests unequivocal even-symmetry. (f) The voltage differential manifests a pronounced even-symmetry characteristic.}
\end{figure*}

  A rectification phenomenon is observed at 4~K under zero applied magnetic field in Device A, using a DC current of 0.6575~mA. This current value was deliberately chosen based on its polarity-dependent response: while a positive current maintains a zero-resistance state, a negative current of the same magnitude induces an abrupt transition to the normal resistive state. This polarity-selective behavior arises from the asymmetry in the critical currents, where $I_{c}^{+} > I_{c}^{-}$, thereby enabling a clear half-wave rectification effect, as demonstrated in Fig. 2(c). Furthermore, the rectification effect persists over an extended measurement duration of 1300 s without degradation, strongly indicating that Joule heating is not responsible for the observed current-direction asymmetry in the superconducting response. This long-term stability reinforces the intrinsic nature of the superconducting diode effect in the studied device, pave the way to the application by using iron-based superconductors.

\section{Discussion}
  Our observations bear resemblance to those reported in early studies,such as PbTaSe$_{2}$\cite{Fengshuo}, FeSe\cite{Utane}, CsV$_{3}$Sb$_{5}$\cite{LeT}, BSCCO\cite{QiS}, where strain induced polarization, thermal gradiant, chiral domain or loop current could induce the zero-field diode effect. To elucidate the underlying mechanisms of the SDE, we devised a series of rigorous exclusion experiments.

Firstly, the stress impact experiment was meticulously devised through the implementation of a specialized electrode configuration, as depicted in Fig.  3(a). During electrode fabrication, reactive ion beam etching was employed to achieve a 27 nm profile, followed by the deposition of a 29 nm Ti/Au film. The distinction lies in the left voltage terminal, featuring a pristine SiO$_2$ surface, whereas the right terminal incorporated a 1.3 $\mu$m recess etched via RIE. Upon sample transfer to the electrodes, intrinsic stress gradients naturally manifested along the edges, thereby enabling a concurrent comparative analysis of stress effects on the SDE. We conducted a comparative analysis of the superconducting diode effect in regions with and without significant applied stress, specifically examining the dependence of the rectification coefficient $\eta$ on the magnetic field. Intriguingly, we observed that the rectification coefficient $\eta$ in the regions without significant stress exhibited a distinct even-symmetric dependence on the magnetic field, a finding that is consistent with the results obtained from the irregularly structured sample in Fig. 3(b). Samely, in the regions with significant applied stress, while the rectification coefficient $\eta$ maintained an even-symmetric characteristic and the rectification coefficient $\eta$ has exhibited 1\% enhancement, indicating the stress can influence the rectification coefficient of SDE in zero-field in Fig. 3(c).

  Secondly, to ascertain the provenance of the zero-field diode phenomenon within our experimental samples, we employed a dry transfer technique coupled with micro-nano fabrication methods to pattern both symmetric and asymmetric architectures upon a singular FTS flake, thereby enabling a systematic investigation into thermal gradient effects in Fig. 4(a). Inspired by asymmetric FeSe reported recently \cite{Utane}, discrepancies in the current terminal contacts of the specimen may establish a persistent thermal gradient, thereby inducing the observed voltage differential($\Delta V_{heat}$), as illustrated in Fig. 4(b). Then, we measured the dependence of the rectification coefficient $\eta$ on the applied out-of-plane magnetic field, as shown in Fig. 4(c). The corresponding raw data are presented in Fig. S6. Notably, the rectification coefficient exhibits an even symmetry with respect to the magnetic field direction, gradually decreasing with increasing field strength and dropping to approximately 0.3\% under a 14~T magnetic field. Voltage differentials under forward and reverse currents of equal magnitude across varying magnetic fields. As quantified in Fig. 4(d) (results on the order of $10^{-5}$  V), the negligible voltage disparity confirmed that Joule heating at the terminals is insufficient to account for the observed SDE. We observed that the dependence of both positive and negative currents($\Delta I_{c}$) on the magnetic field exhibits an even symmetry, whereas $\Delta V_{heat}$ shows no discernible correlation with the magnetic field.  They argue that the Seebeck coefficients independence from the magnetic field aligns with their observations. For irregularly structured samples, we similarly investigated $\Delta V_{heat}$ under forward and reverse currentsthe dependence of the magnetic field. Due to the presence of a significant thermoelectric voltage at the current terminals, both $\Delta I_{c}$ and voltage difference exhibited a distinct even-symmetric dependence on the magnetic field in Fig. 4(e)(f). This observation further excludes the superconducting diode effect originating from thermal gradients. However, this contradicts our findings, which show a suppression of SDE under a magnetic field. Additionally, our sample exhibits a more symmetric shape. Therefore, although the thermoelectric explanation is plausible, it does not fully align with the phenomena we observed.

\begin{figure*}
\includegraphics[width=0.9\textwidth]{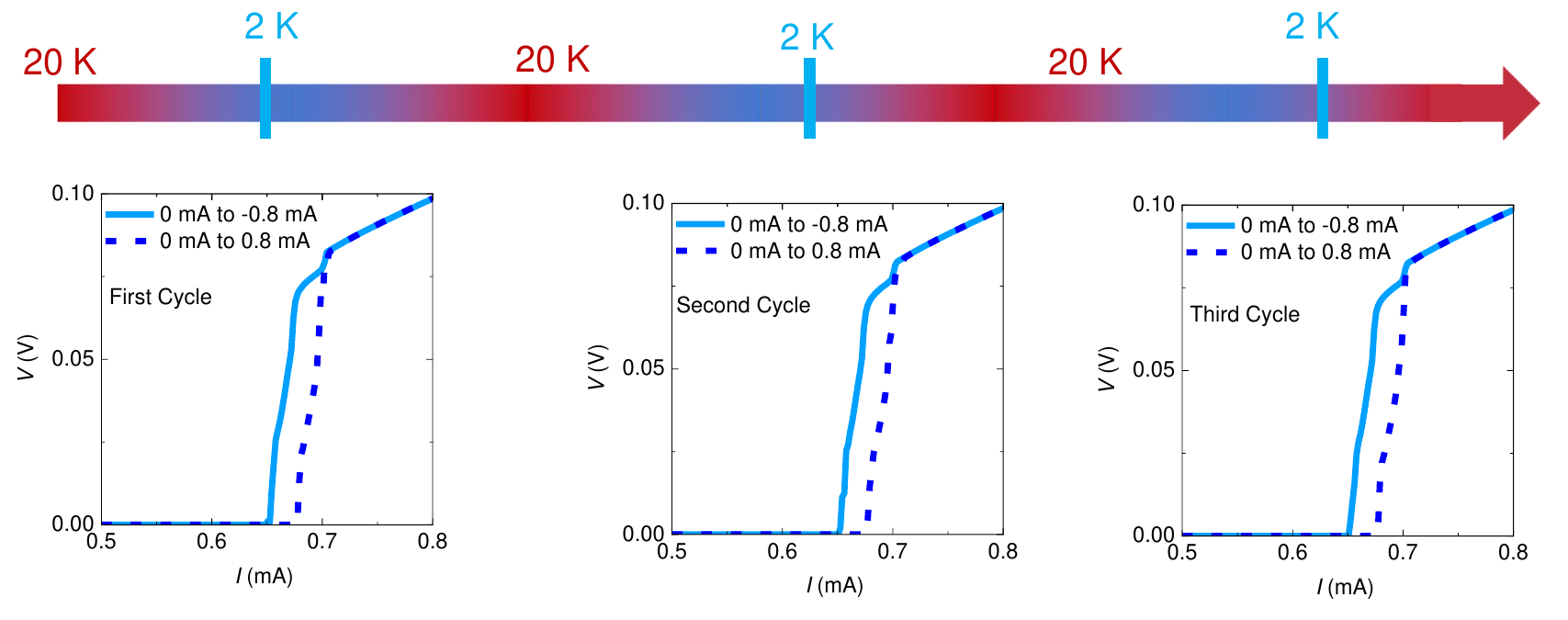} 
\caption{\label{Fig.5} Repeated temperature cycles from superconducting state to normal state were implemented to eliminate potential influences from internal dynamic SC domains.}
\end{figure*}


  Thirdly, in the context of single-crystal FeTe\textsubscript{0.55}Se\textsubscript{0.45}, translational symmetry remains intact, with the crystal momentum \textbf{\emph{k}} is well-defined. This ensures that two electrons with momenta \textbf{\emph{k}} and -\textbf{\emph{k}} are related through a time-reversal symmetry operation. Consequently, a straightforward symmetry argument implies that SDE cannot occur under time-reversal symmetry. We then consider the effect of time-reversal symmetry breaking. Given the topological properties of FeTe\textsubscript{0.55}Se\textsubscript{0.45}, it is conceivable that time-reversal symmetry might be spontaneously broken in its superconducting state. This compound is known to exhibit a bulk s-wave behavior \cite{Hanaguri}, while its surface demonstrates a topological state that also breaks time-reversal symmetry. If the SDE were due to superconductivity with broken time-reversal symmetry, we would expect the sign of \emph{I\textsubscript{c}} to switch between +\emph{B} and -\emph{B}, as observed in valley-polarized trilayer graphene \cite{LinJX}. However, our measurements show that the field dependence of \emph{I\textsubscript{c}} is \emph{B}-even. 
On the other hand, the superconducting domains that with different time-reversal symmetry could affect the SDE coefficient. Based on our initial hypothesis regarding the thermal modulation of the superconducting diode effect, we systematically elevated the sample temperature to its normal state, subsequently restored the oscillating magnetic field to 0 Oe, and then cooled it back to the superconducting state. This procedure was meticulously repeated for three consecutive cycles while monitoring the current-voltage (I-V) transport characteristics. As illustrated in Fig. 5, the I-V properties remained invariant after each thermal cycle, thereby conclusively excluding the presence of SC domains with broken TRS. Because Characteristics of the domains, such as domain asymmetry and inter-domain interaction, are randomly altered by thermal cycling.
Thus, spontaneous breaking of time-reversal symmetry is not the primary cause of the SDE in this system.

  The fourth interpretation proposes that a zero-field SDE does not necessarily rely on broken time-reversal symmetry. Traditionally, it is believed that SDE is forbidden under translational symmetry, as previously discussed. However, introducing electron-electron interactions allows for SDE even without breaking time-reversal symmetry. Theoretical considerations suggest that electron correlation effects lead to a charging effect in the polar direction, resulting in an asymmetric dependence on current direction and, consequently, a zero-field diode effect in the normal state. 
In Fig. S11, we normalized the RT curves of all the measured devices, and red lines represents for the devices with SDE and blue lines represents for that without SDE. Despite the nearly the same superconducting transition, some of the devices reveal an upturn before superconducting transiton while some of the devices show a more metallic states. This could be contributed to the inhomogeneity of the sample. The magnetoresistance changes sign with temperature, suggesting competition between Drude-like transport and Kondo-like scattering. The observed Kondo scattering suggests that the local magnetic moments are likely associated with precipitated interstitial iron \cite{Jinsheng}. We therefore speculate that variations in local strain and electronic correlations could contribute to the device-to-device variation in rectification coefficient. We then suggest that FeTe\textsubscript{0.55}Se\textsubscript{0.45} may inherently fulfill the conditions for this SDE mechanism.

\section{Conclusion}

In conclusion, we investigated the emergence of field-free nonreciprocal transport phenomena in the iron-based superconductor FeTe\textsubscript{0.55}Se\textsubscript{0.45}. Our experimental results demonstrate that the disparity in critical currents is approximately 1.9 \% of the average value at 2 K.  Notably, the second harmonic resistance showed a significant non-zero signal at the superconducting transition, with the signal is polarity depending on the current excitation or thermal conditions. The symmetric response observed under magnetic influence indicates a mechanism distinct from previously reported zero-field superconducting diode effects (SDE). Through meticulously designed experiments, we have systematically ruled out factors such as superconducting dynamic domains, thermal gradients, and geometric configurations, ultimately attributing the observed phenomenon to the influence of localized stress. Stress may serve as the origin of broken inversion symmetry or induced polarization, while strong correlations further amplify nonlinear charge transport and nonreciprocal responses—collectively giving rise to a zero-field SDE. This occurrence of field-free SDE could simplify the integration of superconducting devices and reduce interference, offering a promising pathway for the advancement of superconducting electronics based on iron-based high-temperature superconductors.

\begin{acknowledgments}
This research was supported in part by the Ministry of Science and Technology (MOST) of China (No. 2022YFA1603903), the National Natural Science Foundation of China (Grants No. 62571327, 92565112), Natural Science Foundation of Shanghai (Grants Nos. 25ZR1402374, 25ZR1402368, and 25ZR1401252), the Science and Technology Commission of Shanghai Municipality, the Shanghai Leading Talent Program of Eastern Talent Plan, and the Double First-Class Initiative Fund of ShanghaiTech University.
\end{acknowledgments}


%
\end{document}